\documentclass[onecolumn,showpacs]{revtex4}

\usepackage{graphicx}
\usepackage{dcolumn}
\usepackage{bm}

\input epsf

\begin{document}

\title{\Large  VALIDITY OF THE GENERALIZED SECOND LAW OF THERMODYNAMICS OF THE UNIVERSE BOUNDED
BY THE EVENT HORIZON IN HOLOGRAPHIC DARK ENERGY MODEL}

\author{\bf~Nairwita~Mazumder\footnote{nairwita15@gmail.com}, Subenoy~Chakraborty\footnote{schakraborty@math.jdvu.ac.in}.}

\affiliation{$^1$Department of Mathematics,~Jadavpur
University,~Kolkata-32, India.}

\date{\today}

\begin{abstract}
In this letter, we investigate the validity of the generalized
second law of thermodynamics of the universe bounded by the event
horizon in the holographic dark energy model. The universe is
chosen to be homogeneous and isotropic and the validity of the
first law has been assumed here. The matter in the universe is
taken in the form of non-interacting two fluid system- one
component is the holographic dark energy model and the other
component is in the form of dust.
\end{abstract}


\maketitle

\section{\normalsize\bf{Introduction}}

At present it is strongly believed that our universe is
experiencing an accelerated expansion. The numerous cosmological
observations suggest that the acceleration is driven by a missing
energy density with negative pressure is called as dark energy. An
approach to the problem of dark energy is holographic model[1-13].
The holographic principle states that the no. of degrees of
freedom for a system within a finite region should be finite and
is bounded roughly by the area of its boundary. From the effective
quantum field theory one obtains the holographic energy density
as[14]

$$\rho_{D}=3c^2{M_{p}}^2L^{-2} $$
where $L$ is an IR cut-off in units ${M_{p}}^{2}=1$. Li shows
that[1] if we choose $L$ as the radius of the future event
horizon, we can get the correct equation of state and get the
desired accelerating universe. Also in equation (1) $c~$is any
free dimensionless parameter whose value is determined by
observational data[8,15-19]. However, in the present work we
have taken $c$ to be arbitrary.\\

From the string theory point if view the dimension of space-time
should be more than four. However, Einstein-Hilbert(EH) action is
the most general one in 4-D. For higher dimension general
lagrangian contains Lovelock terms in addition to EH term and we
have second order field equations. The simplest extension is to
take the Gauss-Bonnet(GB)[19-24] term in addition to EH term.
Further GB term is important both from geometrical and physical
point of view. The GB term in the lagrangian is the higher
curvature correction to general relativity and naturally arises as
the next leading order of the $\alpha$ expansion of heterotic
super string theory[25], where $\alpha^{-1}$ measures the string
tension. Geometrically in five dimensions, Einstein-Hilbert action
with GB term gives the most general lagrangian[26] for the second
order field equations, while in four dimensions, the lagrangian
corresponding to EH action is the most general one and GB term is
only the topological one, which does not affect the dynamics[27].\\

In this work we have examined the validity of the generalized
second law of thermodynamics of the universe bounded by the event
horizon assuming the first law both for Einstein gravity as well
as Gauss-Bonnet gravity.\\

\section{\normalsize\bf{Validity of the Generalized Second Law:}}

\subsection{\normalsize\bf{Einstein Gravity}}

In homogeneous and isotropic FRW model, the space-time metric can
be written as

\begin{equation}
ds^{2}=h_{ab}dx^{a}dx^{b}+{\tilde{r}}^{2}d{{\Omega}_{2}}^{2}
\end{equation}

where
$d{\Omega_{2}}^{2}=d\theta^{2}~+~({\sin}^{2}\theta)~d\phi^{2}$,~is
the metric on unit 2- sphere, $\tilde{r}=ar$ is the area radius
and~~$h_{ab}=~diag(-1,\frac{a^{2}}{(1-kr^{2})})$~~with
~$k=0,\pm1$~for flat,closed and open model. The Friedmann
equations are (choosing $8\pi G~=~1$~)

\begin{equation}
H^{2}+\frac{k}{a^{2}}=\frac{\rho_{t}}3
\end{equation}

and

\begin{equation}
\dot{H}-\frac{k}{a^{2}}=~-\frac{1}2 (\rho_{t}+p_{D})
\end{equation}

with~$\rho_{t}=\rho_{m}+\rho_{D}$~. Here $\rho_{D}$~and~$p_{D}$~
correspond to energy density and thermodynamic pressure of the
holographic dark energy model while $\rho_{m}$~is the energy
density corresponding to dust. As the two component matter system
is non interacting so they satisfy the energy conservation
equation separately, i.e.

\begin{equation}
\dot{\rho_{m}}+3H(\rho_{m})=0
\end{equation}

and

\begin{equation}
\dot{\rho_{D}}+3H(\rho_{D}+p_{D})=0
\end{equation}

We shall use the holographic dark energy (DE) model in reference
[1]. As the universe is bounded by the event horizon so the energy
density of the holographic model can be written as

\begin{equation}
\rho_{D}=3c^{2}{R_{E}}^{-2}
\end{equation}

From above using the definition of the cosmological event horizon

\begin{equation}
{R}_{E}=a\int^{\infty}_{a}\frac{da}{Ha^{2}}=\frac{c}{(\sqrt{\Omega_{D}})H}
\end{equation}

where ~$\Omega_{D}=\frac{\rho_{D}}{3H^{2}}$~is the density
parameter corresponding to dark energy. The equation of state of
the dark energy can be written as

\begin{equation}
\rho_{D}=\omega_{D}p_{D}
\end{equation}

where~$\omega_{D}$~is not necessarily a constant.\\

Now the amount of energy crossing the event horizon in time $dt$~
has the expression [28,29]

\begin{equation}
-dE=4\pi{R_{E}}^{3}H(\rho_{t}+p_{D})dt~
\end{equation}

So assuming the validity of the first law of thermodynamics we
have

\begin{equation}
\frac{dS_{E}}{dt}=\frac{4\pi {R_{E}}^3H}{T_{E}}(\rho_{t}+p_{D})
\end{equation}

where $S_{E}$~and~$T_{E}$~are the entropy and temperature of the
event horizon respectively.\\

In order to obtain the variation of the entropy of the fluid
(mentioned earlier) inside the event horizon we use the Gibb's
equation [30,31]

\begin{equation}
T_{E}dS_{I}=dE_{I}+p_{D}dV
\end{equation}

where ~$S_{I}$~ and ~$E_{I}$~ are the entropy and energy of the
matter distribution and for the thermodynamical equilibrium the
temperature of the matter distribution is assumed to be same as
event horizon[32-34] . So starting with
$$E_{I}=\frac{4}3\pi {R_{E}}^{3}\rho_{t}~~and~~V=\frac{4}3\pi
{R_{E}}^{3}~,$$

and using the Friedmann equations the Gibb's equation leads to

\begin{equation}
dS_{I}=\frac{4\pi {R_{E}}^{2}}{T_{E}}(\rho_{t}+p_{D})dR_{E}+
\frac{H{R_{E}}^{3}}{T_{E}}(\dot{H}-\frac{k}{a^{2}})dt
\end{equation}

Now to obtain the change in the radius of the event
horizon~$(R_{E})$~we start with the expression from holographic
dark energy (i.e. equation(7)~) and again using the Friedmann
equations and the conservation equation (5) we get (after
simplification)

\begin{equation}
dR_{E}=\frac{3}2R_{E}H(1+\omega_{D})dt
\end{equation}

Hence using this expression of $dR_{E}$ in (12) we get

\begin{equation}
\frac{dS_{I}}{dt}=\frac{2\pi
{R_{E}}^3}{T_{E}}H(\rho_{t}+p_{D})(3\omega_{D}+1)
\end{equation}

Hence combining (10) and (14) the resulting change of total
entropy is given by

\begin{equation}
\frac{d}{dt}(S_{I}+S_{E})=\frac{6\pi {R_{E}}^3H}{T_{E}}
(\rho_{t}+p_{D})(\omega_{D}+1)
\end{equation}

or equivalently using the deceleration parameter
$q=-1-\frac{\dot{H}}{H^{2}}$ we obtain

\begin{equation}
\frac{d}{dt}(S_{I}+S_{E})=\frac{12\pi {R_{E}}^3H}{T_{E}}
\left[(1+q)H^{2}+\frac{k}{a^{2}}\right](\omega_{D}+1)
\end{equation}

The above results lead to the following conclusions :\\

{\bf I.}~The generalized second law of thermodynamics will be
valid on the event horizon if the holographic dark energy
component individually satisfies the weak energy condition i.e.
$$\rho_{D}+p_{D}=\rho_{D}(1+\omega_{D})>0~.$$
i.e if the holographic dark energy component is not of the phantom
nature then universe as a thermodynamical system with two
non-interacting fluid components(as in the present case) always
obey the second law of thermodynamics. Also from equation (16) the
expression within square bracket will be positive definite if
$k=0,~+1$. For $k=-1$ we must have the inequality
$$\frac{1}{a^{2}H^{2}}<(1+q)$$ for the validity of the generalized second law of
thermodynamics.\\
Further one may note that, in this case the entropy of the event
horizon also increases with time while variation of the matter
entropy with time is not positive definite,but the sum of the
entropies increases with the evolution of the universe. Also the
radius of the event horizon increases with time.\\

{\bf II.}~If we consider the two fluid system as a single fluid
with energy density and pressure
$$\rho_{t}=\rho_{m}+\rho_{D}~,~p_{t}=p_{D}$$ and assume the weak
energy condition for the combined matter system then
$$\rho_{t}+p_{t}>0$$ i.e. $$\rho_{m}+\rho_{D}(1+\omega_{D})>0~.$$
Clearly the above inequality does not guarantee $$(1+\omega_{D})>0$$
 i.e the weak energy condition for the holographic dark energy.
  Consequently from equations (15) and (16), the validity of the second
 law of thermodynamics is not definite. More over from equation(16) the
 expression within square bracket may be positive even if $q$ is negative i.e.
  if  $$q>-1-\frac{k}{a^{2}H^{2}}~.$$ Therefore the generalized second law may be
valid both for a accelerating and decelerating phase of the
universe.

\subsection{\normalsize\bf{Gauss-Bonnet Gravity:}}

Due to complicated form of the field equations we consider only
the flat FRW model in GB theory. In $(n+1)-$dimensional flat FRW
model the modified Einstein equations in GB gravity are [29]
(choosing $\frac{16\pi G}{n-1}=1~$)

\begin{equation}
H^{2}(1+\tilde \alpha H^{2})=\frac{\rho_{t}}n
\end{equation}

and

\begin{equation}
\dot{H}(1+2\tilde \alpha H^{2})=-\frac{\rho_{t}+p_{D}}2
\end{equation}

with the conservation equations

\begin{equation}
\dot{\rho_{m}}+nH(\rho_{m})=0
\end{equation}

and

\begin{equation}
\dot{\rho_{D}}+nH(\rho_{D}+p_{D})=0
\end{equation}

As, before we have the same expression for the radius of the
event horizon considering the holographic dark energy model [1]
i.e.$${R}_{E}=\frac{c}{(\sqrt{\Omega_{D}})H}$$ where in GB theory
density parameter has the expression

\begin{equation}
\Omega_{D}=\frac{\rho_{D}}{nH^{2}(1+\tilde \alpha H^{2})}
\end{equation}

Thus using the above modified Einstein field equation a small
variation of the radius of the event horizon is given by

\begin{equation}
dR_{E}=HR_{E}\left[\frac{n}2(1+\omega_{D})+\frac{\tilde \alpha
\dot{H}}{(1+\tilde \alpha H^{2})}\right]dt
\end{equation}

As the amount of energy crossing the horizon in time $dt$ does
not depend on any particular gravity theory so from the first law
of thermodynamics the time variation of the entropy of the event
horizon is given by

\begin{equation}
\frac{dS_{E}}{dt}=\frac{n\Omega_{n}
{R_{E}}^nH}{T_{E}}(\rho_{t}+p_{D})
\end{equation}

where~$\Omega_{n}=\frac{{\pi}^{\frac{n}2}}{\Gamma(\frac{n}2+1)}$~is
the volume of an n-dimensional unit ball.\\

Again using Gibb's law (11) and the variation of $R_{E}$ (given
by equation (21)) and proceeding as in Einstein gravity the rate
of change of entropy (with respect to time) of the matter bounded
by the event horizon can be expressed as

\begin{equation}
\frac{dS_{I}}{dt}=\frac{n\Omega_{n}{R_{E}}^{n}H}{T_{E}}(\rho_{t}+p_{D})
\left[\frac{n}2(1+\omega_{D})~+~\frac{\tilde \alpha
\dot{H}}{(1+\tilde \alpha H^{2})}-1\right]
\end{equation}

where we have to use the modified Einstein field equations
(17)-(18) and the conservation equation (19), to derive the above
expression.\\

Thus combining (22) and (23) the time variation of the total
entropy is given by

\begin{equation}
\frac{d}{dt}(S_{I}+S_{E})=\frac{n\Omega_{n}{R_{E}}^{n}H(\rho_{t}+p_{D})}{2T_{E}}
\left[n(1+\omega_{D})+\frac{2\tilde \alpha \dot{H}}{(1+\tilde
\alpha H^{2})}\right]
\end{equation}

or equivalently using deceleration parameter

\begin{equation}
\frac{d}{dt}(S_{I}+S_{E})=\frac{n\Omega_{n}{R_{E}}^{n}H^{3}}{T_{E}}(1+q)(1+2\tilde
\alpha H^{2}) \left[n(1+\omega_{D})-\frac{2\tilde \alpha
(1+q)H^{2}}{(1+\tilde \alpha H^{2})}\right]
\end{equation}

In this case we cannot make any definite conclusion regarding
validity of the second law of thermodynamics. The term within the
square bracket in the r.h.s. of the equation (24) (or (25))
suggests that the equation of state for the holographic dark
energy should be restricted by a complicated expression involving
$H$ and $\dot{H}$. Finally note that, as $\alpha \rightarrow 0$
we get back our earlier results in Einstein gravity.\\

\section{\normalsize\bf{Conclusions:}}

In the present work we examine the validity of the generalized
second law of thermodynamics on the event horizon assuming the
validity of the first law of thermodynamics both in Einstein
gravity as well as in Gauss-Bonnet gravity. We consider the
universe as a thermodynamical system and is filled with
non-interacting two fluids. Here one component has been considered
 as the holographic dark energy and other is in the form of
 dust. One may note that we have not used the blackhole entropy as the entropy
 of the event horizon. In the first section we have shown that if the weak energy
 condition is satisfied by the holographic dark energy component
 on the event horizon then the generalized second law is valid
 there provided the first law is valid on the event horizon. It is
 to be noted that a similar study was done by Horvat[35] for universe
 filled with holographic dark energy. So our work may be considered as
 a generalization of his work with non-interacting two fluid system. The
 next section deals with the flat FRW model in Gauss-Bonnet
 theory. Here also considering the holographic dark energy
 scenario in the frame work of Gauss-Bonnet theory we have studied
 the generalized second law of thermodynamics on the event horizon
 assuming the validity of the first law. But due to complicated
 form of the expressions in Gauss Bonnet gravity we can not make
 any definite conclusion as in the previous case. The criteria of
 the validity of the first law of thermodynamics on the event
 horizon will be interesting and subsequently we shall study it.\\

{\bf Acknowledgement:}\\

The paper is done during a visit to IUCAA, Pune, India.The authors
are thankful to IUCAA for warm hospitality and facility of doing
research works.\\

{\bf References:}\\
\\

$[1]$ M. Li , {\it Phys. Lett. B} {\bf 603} 01 (2004);\\\\
$[2]$ M. R. Setare  and S. Shafei  , {\it JCAP } {\bf 0609} 011
(2006) arXiv: gr-qc/0606103 ;\\\\
$[3]$ B. Hu and Y. Ling \it{Phys. Rev. D} {\bf 73} 123510
(2006).\\\\
$[4]$ B. Wang , Y. Gong , E. Abdalla, \it{Phys. Lett. B} {\bf 624} 141
(2005).\\\\
$[5]$ M. Ito , {\it Europhys. Lett. } {\bf 71} 712 (2005);\\\\
$[6]$ S. Nojiri , S. Odintsov , {\it Gen. Rel. Grav.} {\bf 38} 1285
(2006);\\\\
$[7]$ E.N. Saridakis , {\it Phys. Lett. B} {\bf 660} 138
(2008);\\\\
$[8]$ Q.G. Huang and M. Li , {\it JCAP} {\bf 0408} 013 (2004);\\\\
$[9]$ X. Zhang , {\it IJMPD } {\bf 14} 1597 (2005);\\\\
$[10]$ D. Pavon and W. Zimdahl , {\it Phys. Lett. B} {\bf 628} 206
(2005);\\\\
$[11]$ H. kim , H.W. Lee and Y.S. Myung , {\it Phys. Lett. B} {\bf 632} 605
(2006);\\\\
$[12]$ S.D. Hsu , {\it Phys. Lett. B} {\bf 594} 01 (2004);\\\\
$[13]$ R. Horvat , {\it Phys. Rev. D} {\bf 70} 087301 (2004);\\\\
$[14]$ A.G. Cohen , D.B. Kaplan and A.E. Nelson , {\it Phys. Rev. Lett.} {\bf 82} 4971
(1999);\\\\
$[15]$ Z. Chang, F-Q. Wu and X. Zhang, Phys. Lett. B 633, 14
(2006) [arXiv:astroph/ 0509531].\\\\
$[16]$ H. C. Kao, W. L. Lee and F. L. Lin, astro-ph/0501487; X.
Zhang and F-Q. Wu, Phys. Rev. D 72, 043524 (2005)
[arXiv:astro-ph/0506310]; X. Zhang and F-Q. Wu, Phys. Rev. D 76,
023502 (2007) [arXiv:astro-ph/0701405].\\\\
$[17]$ Q. Wu, Y. Gong, A. Wang and J. S. Alcaniz,
[arXiv:astro-ph/0705.1006]; Y-Z. Ma and Y. Gong,
[arXiv:astro-ph/0711.1641].\\\\
$[18]$ J. Shen, B. Wang, E. Abdalla and R. K. Su,
[arXiv:hep-th/0412227].\\\\
$[19]$ E.N. Saridakis and M.R. Setare , {\it Phys. Lett. B} {\bf
670} 01 (2008);\\\\
$[20]$ E.N. Saridakis , {\it Phys. Lett. B} {\bf 661} 335
(2008);\\\\
$[21]$ C. Charmouis and J.F. Dufaux , {\it CQG} {\bf 19} 4671
(2002);\\\\
$[22]$ J.E. Kim , B. Kyae  and H.M. Lee , {\it Nucl. Phys. B} {\bf
582} 296 (2000);\\\\
$[23]$ E. Gravanis  and S. Willison , {\it Phys. Lett. B } {\bf
562} 118
(2003);\\\\
$[24]$ R.G. Cai , H.S. Zhang  and A. Wang , {\it Commun Theor.
Phys. } {\bf 44} 948 (2005);\\\\
$[25]$ P. Candelas , G.T. Horowitz , A. Strominger and E. Witten
{\it Nucl. Phys. B} {\bf 258} 46 (1985); M.B.~ Greens,~~ J.H.
Schwarz and E. Witten ,"Super String Theory", {\it Cambridge
University Press} (1987); ~~J. ~Polchinski , "String Theory", {\it
Cambridge University Press, Cambridge}, (1998).\\\\
$[26]$ C. Lanczos ,{\it Ann. Math } {\bf 39} 842 (1938).\\\\
$[27]$ M. Sami and N. Dadhich , {\it TSPU Vestink } {\bf 44N7 } 25
(2004) (arXiv: hep-th /0405016).\\\\
$[28]$R.S. Bousso, \it{Phys. Rev. D} {\bf 71} 064024 (2005).\\\\
$[29]$ R. G. Cai and S. P. Kim, {\it JHEP} {\bf 02} 050
(2005).\\\\
$[30]$ G. Izquierdo and D. Pavon, {\it Phys. Lett. B} {\bf 633}
420 (2006).\\\\
$[31]$ B. Wang, Y. Gong, E. Abdalla, \it{Phys. Rev. D} {\bf 74}
083520 (2006).\\\\
$[32]$ E N. Saridakis, P.F. Gonz´alez-D´ýaz and C.L. Sig¨uenza ,
arXiv: 0901.1213/astro-ph;\\\\
$[33]$ P.F. Gonz´alez-D´ýaz and C.L. Sig¨uenza ,{\it Nucl. Phys.
B} {\bf 697} 363 (2004);\\\\
$[34]$ S. H. Pereira and J.A.S. Lima , {\it Phys. Lett. B} {\bf 669} 266 (2008);\\\\
$[35]$ R. Horvat , {\it Phys. Lett. B} {\bf 664} 201 (2008);\\\\

\end{document}